\documentclass[sigchi-a, nonacm=true, authorversion = true]{acmart}
\usepackage{booktabs} 
\usepackage{ccicons}  
\usepackage{pifont}
\usepackage[flushleft]{threeparttable}
\def\BibTeX{{\rm B\kern-.05em{\sc i\kern-.025em b}\kern-.08emT\kern-.1667em\lower.7ex\hbox{E}\kern-.125emX}}
    
\setcopyright{rightsretained}

\copyrightyear{2019}
\acmYear{2019}
\setcopyright{rightsretained}
\acmConference[CHI 2019] {CHI Conference on Human Factors in Computing Systems Proceedings}{May 4--9, 2019}{Glasgow, Scotland UK}

\begin{document}
\title{Finding Design Opportunities for Smartness in Consumer Packaged Goods}

%

\author{Gustavo Berumen}
\affiliation{
  \institution{Mixed Reality Lab \& Horizon CDT\newline
  University of Nottingham}
  \city{Nottingham}
  \postcode{NG8 1BB}
  \country{UK}}
\email{gustavo.berumen@nottingham.ac.uk}

\author{Joel E. Fischer}
\affiliation{
  \institution{Mixed Reality Lab\newline
  University of Nottingham}
    \city{Nottingham}
  \postcode{NG8 1BB}
  \country{UK}}
\email{joel.fischer@nottingham.ac.uk}

\author{Anthony Brown}
\affiliation{
  \institution{Horizon Digital Economy Research Institute\newline
  University of Nottingham}
  \city{Nottingham}
  \postcode{NG8 2TU}
  \country{UK}}
\email{anthony.brown@nottingham.ac.uk}

\author{Martin Baumers}
\affiliation{
  \institution{Centre for Additive Manufacturing\newline
  University of Nottingham}
  \city{Nottingham}
  \postcode{NG7 2GX}
  \country{UK}}
\email{martin.baumers@nottingham.ac.uk}

%
\renewcommand{\shortauthors}{G. Berumen et al.}

\begin{abstract}

This study attempts to understand the use of Consumer Packaged Goods (CPG) in practice to obtain insights to develop design interventions that bring the CPGs into the Internet of Things. Our ultimate aim is to equip CPGs with a layer of smartness so that CPGs could collect information about their use and provide extra services and functionalities. With a practice perspective we developed an assemblage of methods to analyze and represent how people use CPGs. We chose cooking as our practice case and use an autoethnographic data sample to demonstrate the application of our methods. Despite the early stage of our study, our methods provide ways to get an understanding of how CPGs are used in practice and an opening to establish opportunities for design interventions.

\begin{margintable}
\textit{2019 Copyright held by the author(s)}
\end{margintable}
\end{abstract}

%
%


%

\settopmatter{printacmref = false}

%

\maketitle

\section{Introduction}

Consumer packaged goods (CPG) are low-cost products that experience depletion upon each use ~\cite{borden1964concept}. Packaged food, beverages, and toiletries are examples of CPGs. There is growing industry interest in moving CPGs towards the Internet of Things ~\cite{lee2013internet}; so that CPGs could collect information about their use and provide extra functionalities such as helping to reduce waste, supporting proper product usage, and enhancing the experience of using the products, among others. 

\begin{marginfigure}
    \includegraphics[width=\marginparwidth]{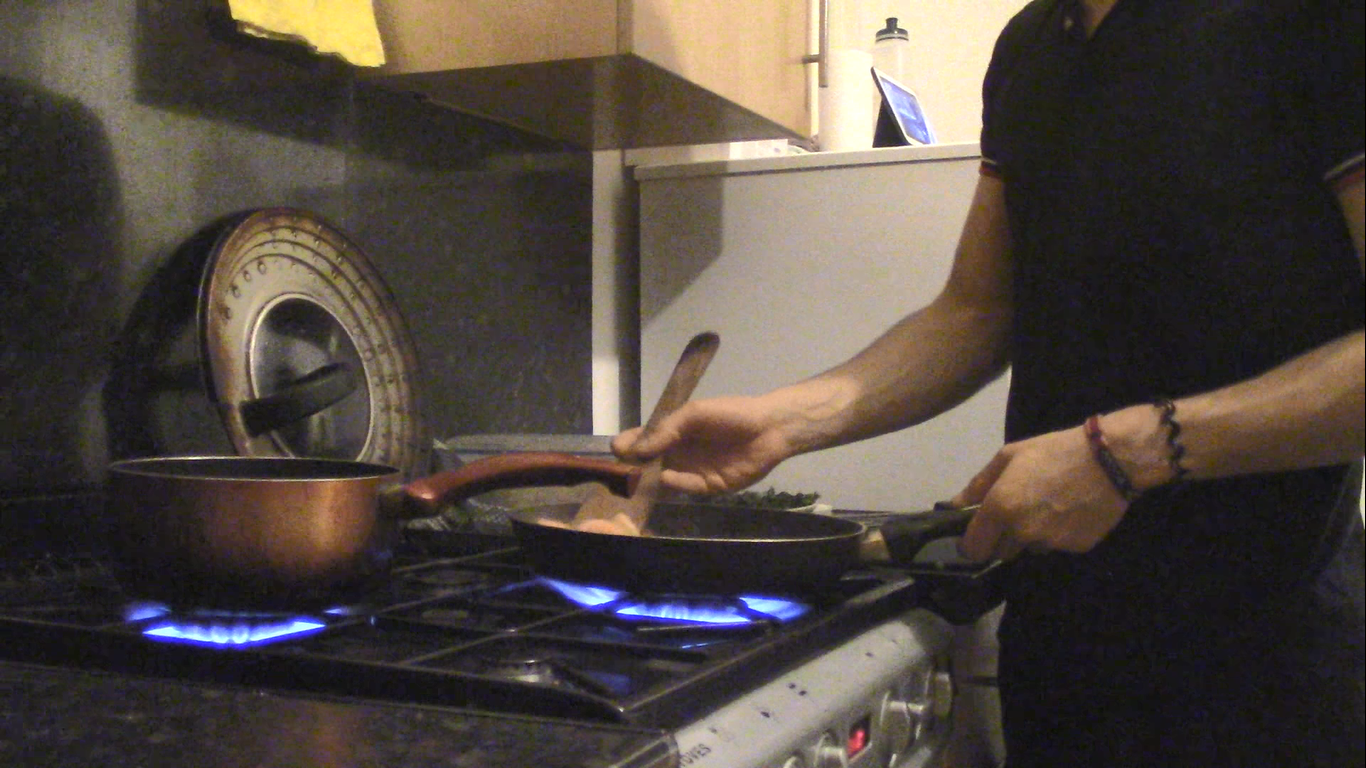}\Description{A sample image of cooking.}
    \caption{The participant cooking a meal.}
    \label{fig:cooking}
\end{marginfigure}

Until now there has not been a structured attempt to develop smart versions CPGs that fit into the practices (a specific ways of conducting a routine \cite{reckwitz2002toward}) they are part of. Current smart CPGs are designed for a specific CPG and for specific situations of use (see examples ~\cite{schneider2008smart, ratnakar2007smart, stylus_2016}). We believe that a practice perspective \cite{kuutti2014turn} is useful to first  understand how CPGs are used, and then based on that understanding develop design interventions ~\cite{crabtree2012doing}. 

Here we aim to investigate whether it is possible to gain insights into how to develop design interventions for CPGs by understanding the use of CPGs in practice. For this purpose we develop a variety of methods, inspired by previous research, to represent the usage of CPGs. We took cooking as our research case an autoethnographic data sample ~\cite{ellis2011autoethnography} to demonstrate the application of our methods. 

\section{Methods and data sample}

\begin{margintable}
 \caption{Items Usage Counting: the tree most and least used CPGs and utensils.}
 \label{tab:table1}
 \begin{threeparttable}
\begin{tabular}{l l l | l l l}
\hline
    & CPGs & {\textit{f }} & &Utensils & {\textit{f }} \\
\hline
1   &water      &7      &1      &palette        &13\\ 
2   &fish       &5      &2      &lid            &7\\ 
3   &greens     &4      &3      &pan            &7\\ 
    &           &       &       &               &\\ 
17  &parley     &1      &11     &google home    &2\\ 
18  &peas       &1      &12     &plate          &2\\ 
19  &salt       &1      &13     &knife          &1\\ 
\bottomrule
\end{tabular}
\begin{tablenotes}
    \item \textit{Notes. } f = frequency of item usage (number of times item was used during the cooking session; Total number of CPGs = 19; Total number of utensils = 13.
\end{tablenotes}
\end{threeparttable}
\end{margintable}

\subsection{Data Sample}

An author of this paper recorded a video of himself cooking a meal (a vegetables and fish dish). The video captured most of the participant's activities from cooking preparation until serving the meal (See Figure  ~\ref{fig:cooking}). 

\subsection{Methods}

\subsubsection{CPGs Inventory}

The first step to understand CPGs' use is to know what CPGs are available in a person's kitchen. Here, we created an inventory by recording an additional video of the CPGs in the participant's kitchen. We included in the inventory CPGs that belong exclusively to the participant (personal) and CPGs that any member of the participant's house could use (shared). In addition, to have an idea of the time that an item could be available, if not consumed, we included a rough estimate of the expected shelf life of CPGs: short (< 2 weeks) and long (> 2 weeks). 

\subsubsection{Item Usage Counting}

To attempt to know what and how relevant CPGs were for the cooking session, we noted the CPGs used in the cooking session and the number of times that they were used, inspired in part by ~\cite{crabtree2016day}. We also include utensils in this counting, because they are closely associated to the use of CPGs. We considered only when items (CPGs and utensils) were used to accomplish something. For example, when the participant grabbed a bottle of salt to add it to food, an usage of the salt was recorded, however, when the bottle of salt was moved to reach another spice, no usage was recorded.

\begin{margintable}
\caption{CPGs' Affordances.}
\label{tab:table2}
\begin{threeparttable}
\begin{tabular}{l l}
\hline
CPGs                 & Affordances\\
\hline
greens      & o,a,h,ct,p,s,d\\
water       & a,cl              \\
kale        & o,a,h,p,s    \\
fish        & o,a,h,p,s,d \\
oil         & o,c,a, h,s     \\
peas        & o,a,h,p       \\
red lentils & a,h,p          \\
squash     & o,a,h,p,s    \\
parsley     & o,a,s           \\
black pepper   & o,c,a,s           \\
butter      & o,c,a,h,ct        \\
soap        & a,cl              \\
sponge      & cl                \\
cloth       & cl                \\
salt        & o,a,s             \\
chilli powder   & o,a,s             \\
cajun powder    & o,a,s             \\
basil dried & o,c,a,s           \\
milk        & o,c,a,s           \\
\bottomrule
\end{tabular}
\begin{tablenotes}
    \item \textit{Notes. } 
    o = opening, c = closing, a = adding, h = heating, ct = cutting, p = positioning, cl = cleaning, s = storing, and d = disposing of. 
\end{tablenotes}
\end{threeparttable}
\end{margintable}

\subsubsection{CPGs' Affordances}

To get an understanding of how CPGs are used in a cooking session, we looked for CPGs' affordances. An affordance is the relationship between the properties of an object and the capabilities of a person that determine how the object could be used  ~\cite{norman2013design}. In other words, the actions that a person could accomplish using an object. We noted the affordances that each of the CPGs used in the cooking session by examining the cooking session video. 

\subsubsection{Consumption Estimation}

To have an estimate of whether the CPGs could be used in more than one cooking session, we made an evaluation of whether the CPGs were used or used up. The estimates were made by the researcher through a visual inspection of the videos and his knowledge of the cooking session. An example of estimation is a bag of frozen green; if half of the content of the bag was consumed, but still, product remains in the bag which it was considered that was used. 

\subsubsection{Network Visualizations}

CPGs are not used in isolation, but rather in combination with other CPGs and utensils. To understand how people accomplish the work of cooking using CPGs and utensils, we represented visually all the relevant interactions that those items have between them, using Ognyanova source code ~\cite{ognyanova_2018}. As an illustrative example, for an onion that was sliced, we may have noted the following interactions: onion-chopping board, knife-onion, knife-chopping board, etc.

\subsubsection{Categorization of Specific Interactions}

To capture situations that poses difficulties that the participant has to solve (trouble) or situations in which a creative or insightful technique helped to accomplish an activity (opportunity). We examine the videos to find and classify the interactions into those two categories according to our own subjective criteria. A trouble interaction was having to remove an extra amount of an ingredient while an example of an opportunity interaction was using an empty package as temporary rubbish. 

\section{Demonstrations}

\subsubsection{CPGs Inventory}

The participant had a total of 156 available CPGs, of which 81 (51.9\%) were personal and 75 (40.1\%) were shared. Of the 156 total CPGs, 24 (15.4\%) were perishable and 132 (84.6\%) were non-perishable. The participant used 19 CPGs in the cooking session; 12 (63.2\%) personal and 7 (36.8\%) shared. Out of those 19 CPGs, 7 were perishable (36.8\%), and 12 non-perishable (63.2\%). The results suggest that the participant used a greater percentage of shared and perishable CPGs than their percentage in regards to the total number of CPGs.

\subsubsection{Item Usage Counting}

The participant used 19 unique CPGs in 46 occasions with a minimum of 1 and maximum of 7 interactions (mean = 2.42, SD = 1.66). The most used CPGs were water, fish, and greens and red lentils. The participant interact with 13 unique utensils in 60  occasions with a minimum of 1 and maximum of 13 interactions (mean =4.61, SD =2.97). The most used utensils were a palette, lid, and pan. The results suggest that both CPGs and utensils are used multiple times in a cooking session and that utensils are used more times than CPG. 

\begin{margintable}
\caption{CPGs Consumption Estimation.}
\label{tab:table3}
\begin{threeparttable}
\begin{tabular}{l l}
\hline
CPGs                 & portion consumed\\
\hline
greens              & 1/1 (up)  \\
water               & NA        \\
kale                & 1/3       \\
fish                & 2/3       \\
oil                 & dash      \\
peas                & 1/4 (up)  \\
red lentils         & 1/4 (up)  \\
butternut squash    & 1/6       \\
parsley             & pinch     \\
black pepper        & pinch     \\
butter              & pinch     \\
soap                & dash      \\
sponge              & wear      \\
cloth               & wear      \\
salt                & pinch     \\
chilli powder       & pinch     \\
cajun powder        & pinch     \\
basil dried         & pinch     \\
milk                & dash      \\
\bottomrule
\end{tabular}
\begin{tablenotes}
    \item \textit{Notes. } portion consumed in relation to a CPG package; up = used up; Items arranged from top to bottom according to the order they were used. 
\end{tablenotes}
\end{threeparttable}
\end{margintable}

\subsubsection{CPGs' Affordances}

We found 9  CPGs' affordances: opening, closing, adding, heating, cutting, positioning, cleaning, storing, and disposing of. The most common affordance was "adding" and the least were "disposing of" and "cutting" with occurrences in 17 and 2 CPGs respectively (Table~\ref{tab:table2}). "Greens" had the most affordances (7) and "sponge" and "cloth" had the least (2) . Interestingly, spices (salt, chili powder, cajun, basil, and black pepper) have similar affordances with opening, closing and storing. The results suggests alike CPGs share many affordances, indicating that similar CPGs may be used in a similar manner. 

\subsubsection{Consumption Estimation}

We found that 3 of the 19 CPGs were used up, those being greens, peas, and red lentils (Table~\ref{tab:table3}). Only 5 CPGs had a considerable used that is a relevant proportion of its total. For instance, a third of a kale package was consumed. For the other 13 CPGs, the consumption was a negligible part, for instance, a pinch of salt. The results suggests that most CPGs could be used more than one time, which is particularly true for non-perishable CPGs.

\subsubsection{Network Visualizations}

\begin{figure*}
  \includegraphics[scale = 0.68]{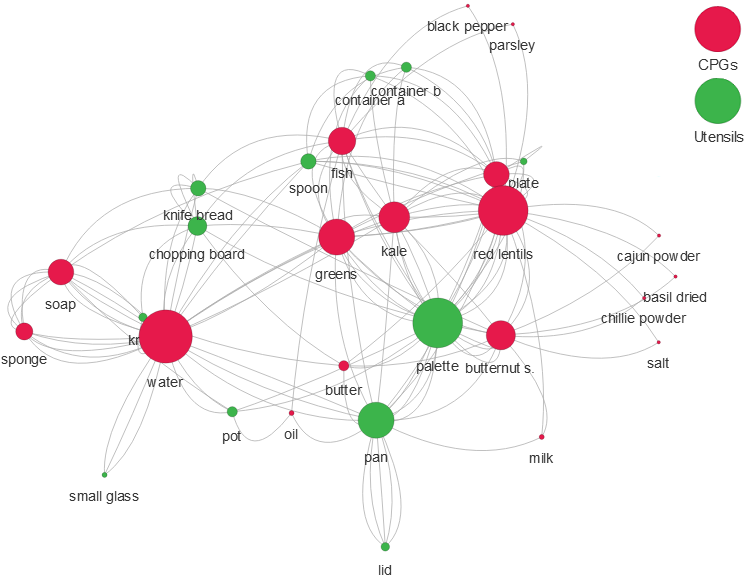}
  \caption{Network visualization that represents the interactions between CPGs and utensils during a cooking session.}~\label{fig:network}
  \Description{A network visualization image that represent the interaction between CPGs and utensils}
\end{figure*}

The size of a node represents the number of interactions it had with other items and the edges represent interactions between nodes (See Figure  ~\ref{fig:network}). The proximity of nodes' centers inversely correlates with the number of interactions between them. The visualization gives a sense for the way items were used during a given cooking session; on the left side, the spices form a cluster that connects to food (lentils and squash). We could infer that those spices were used to season food. There are nodes that have a prominent size as they are connected to a large number of items. Palette is a central node and we could infer that it had a central role in the cooking process. 

\subsubsection{Categorization of Specific Interactions}

We found a total of 9 interactions in the cooking session of those 3 represent trouble interactions and 6 opportunity interactions (Table~\ref{tab:table4}). Next, we provide a detailed description and a digital intervention idea for one trouble and one opportunity interactions:

\textbf{\textit{Fixing meal (as an opportunity)}} This refers to all actions related to improving the taste of food. Examples include seasoning the food, add extra ingredients or changing the cooking method.

\textit{Intervention idea: } CPGs could provide support to users about how to season food. While a person is cooking a meal, CPGs could indicate that they go well with a food.

\textbf{\textit{Unorthodox action (as a trouble)}} Actions on which the person makes an unorthodox use of CPGs or utensils. Using a knife to get a portion of mayonnaise from a jar is an example of this. 

\textit{Intervention idea: } CPGs could take advantage of the hidden unorthodox use of their products. CPGs could be re-designed to better adapt to the ways that people use. 

The classification of the interactions could represent areas for design opportunities as could give designers and researcher situations worth to explore.

\section{Discussion}

Our proposed methods suggest that it is possible to get an understanding of how CPGs are used by focusing on different perspectives and granularity of CPGs usages. Such understanding could inspire different interventions, ranging from general (identifying available CPGs in the kitchen) to specific (revealing the formation of clusters of CPGs). An example of an intervention could be the development of a system that recommends recipe meals based on the available CPGs at home. This could be accomplish by equipping CPGs with low cost sensors that allows them to communicate with a system at home to indicate their presence.

\begin{margintable}
\caption{Trouble and Opportunity Interactions: short description, frequency and trouble to opportunity ratio}
\label{tab:table4}
\begin{threeparttable}
\begin{tabular}{l l l}
\hline
    & Category                                  & \textit{f }(T:O)  \\   \hline
1   & CPGs and utensils placing                 & 2(0:2)          \\
    & {\footnotesize\textit{positioning CPGs and/or utensils}}\\
2   & fixing meal                               & 2(1:1)     \\
    & {\footnotesize\textit{actions to improve food taste}}\\
3  & unhygienic action                         & 2(2:0)     \\
    & {\footnotesize\textit{not being clean enough}}\\ 
4   & extra activities                          & 1(0:1)     \\
    & {\footnotesize\textit{activities carried out while waiting}}\\
5   & unorthodox action                         & 1(0:1)     \\
    & {\footnotesize\textit{actions that deviate from the norm}}\\
6  & use of technology                         & 1(0:1)      \\
    & {\footnotesize\textit{use of devices with computational power}}\\
\bottomrule
\end{tabular}
\begin{tablenotes}
    \item \textit{Notes. } f = frequency of category out of a total of 9 interactions; T:O = trouble to opportunities ratios. 
\end{tablenotes}
\end{threeparttable}
\end{margintable}

The main limitation of our study is the lack of implementations. Although we used a data sample as a way to demonstrate our results, we have not applied our methods in a large number of people, and as now the potential of our methods are only speculations. The results and the applications of our methods may vary with a large number of people and with a longer implementation. We could expect, for instance, our way of counting the interactions between CPGs and utensils to be very time consuming and that may restrict the application in a large number of people.

To summarize, the methods proposed in this paper appear to represent in a fairly acceptable manner how CPGs are used and provide insights for design interventions. As with any attempt to represent the real world, those methods may focus on as well as exclude some information. We may argue for the implementation of the methods in a large number of people. Our immediate next steps will involve both the inclusion of more participants and working towards refining our analysis methods.\

\begin{acks}
This work was supported by the Horizon Centre for Doctoral Training at the University of Nottingham (UKRI Grant No. EP/P510592/1) and by Unilever UK Ltd.
\end{acks}

\bibliographystyle{ACM-Reference-Format}
\bibliography{sample-base}


\begin{thebibliography}{12}


\ifx \showCODEN    \undefined \def \showCODEN     #1{\unskip}     \fi
\ifx \showDOI      \undefined \def \showDOI       #1{#1}\fi
\ifx \showISBNx    \undefined \def \showISBNx     #1{\unskip}     \fi
\ifx \showISBNxiii \undefined \def \showISBNxiii  #1{\unskip}     \fi
\ifx \showISSN     \undefined \def \showISSN      #1{\unskip}     \fi
\ifx \showLCCN     \undefined \def \showLCCN      #1{\unskip}     \fi
\ifx \shownote     \undefined \def \shownote      #1{#1}          \fi
\ifx \showarticletitle \undefined \def \showarticletitle #1{#1}   \fi
\ifx \showURL      \undefined \def \showURL       {\relax}        \fi
\providecommand\bibfield[2]{#2}
\providecommand\bibinfo[2]{#2}
\providecommand\natexlab[1]{#1}
\providecommand\showeprint[2][]{arXiv:#2}

\bibitem[\protect\citeauthoryear{Baron and Dorfer}{Baron and Dorfer}{2016}]%
        {stylus_2016}
\bibfield{author}{\bibinfo{person}{Katie Baron} {and} \bibinfo{person}{Stefanie
  Dorfer}.} \bibinfo{year}{2016}\natexlab{}.
\newblock \bibinfo{title}{Smart Packaging: Absolut's IoT Innovation Lab}.
\newblock
\newblock
\newblock
\shownote{\url{https://www.stylus.com/lcrtjv/}.}


\bibitem[\protect\citeauthoryear{Borden}{Borden}{1964}]%
        {borden1964concept}
\bibfield{author}{\bibinfo{person}{Neil~H Borden}.}
  \bibinfo{year}{1964}\natexlab{}.
\newblock \showarticletitle{The concept of the marketing mix}.
\newblock \bibinfo{journal}{\emph{Journal of advertising research}}
  \bibinfo{volume}{4}, \bibinfo{number}{2} (\bibinfo{year}{1964}),
  \bibinfo{pages}{2--7}.
\newblock


\bibitem[\protect\citeauthoryear{Crabtree, Rouncefield, and Tolmie}{Crabtree
  et~al\mbox{.}}{2012}]%
        {crabtree2012doing}
\bibfield{author}{\bibinfo{person}{Andrew Crabtree}, \bibinfo{person}{Mark
  Rouncefield}, {and} \bibinfo{person}{Peter Tolmie}.}
  \bibinfo{year}{2012}\natexlab{}.
\newblock \bibinfo{booktitle}{\emph{Doing design ethnography}}.
\newblock \bibinfo{publisher}{Springer}.
\newblock


\bibitem[\protect\citeauthoryear{Crabtree and Tolmie}{Crabtree and
  Tolmie}{2016}]%
        {crabtree2016day}
\bibfield{author}{\bibinfo{person}{Andy Crabtree} {and} \bibinfo{person}{Peter
  Tolmie}.} \bibinfo{year}{2016}\natexlab{}.
\newblock \showarticletitle{A Day in the Life of Things in the Home}. In
  \bibinfo{booktitle}{\emph{Proceedings of the 19th ACM Conference on
  Computer-Supported Cooperative Work \& Social Computing}}. ACM,
  \bibinfo{pages}{1738--1750}.
\newblock


\bibitem[\protect\citeauthoryear{Ellis, Adams, and Bochner}{Ellis
  et~al\mbox{.}}{2011}]%
        {ellis2011autoethnography}
\bibfield{author}{\bibinfo{person}{Carolyn Ellis}, \bibinfo{person}{Tony~E
  Adams}, {and} \bibinfo{person}{Arthur~P Bochner}.}
  \bibinfo{year}{2011}\natexlab{}.
\newblock \showarticletitle{Autoethnography: an overview}.
\newblock \bibinfo{journal}{\emph{Historical Social Research/Historische
  Sozialforschung}} (\bibinfo{year}{2011}), \bibinfo{pages}{273--290}.
\newblock


\bibitem[\protect\citeauthoryear{Kuutti and Bannon}{Kuutti and Bannon}{2014}]%
        {kuutti2014turn}
\bibfield{author}{\bibinfo{person}{Kari Kuutti} {and} \bibinfo{person}{Liam~J
  Bannon}.} \bibinfo{year}{2014}\natexlab{}.
\newblock \showarticletitle{The turn to practice in HCI: towards a research
  agenda}. In \bibinfo{booktitle}{\emph{Proceedings of the 32nd annual ACM
  conference on Human factors in computing systems}}. ACM,
  \bibinfo{pages}{3543--3552}.
\newblock


\bibitem[\protect\citeauthoryear{Lee, Crespi, Choi, and Boussard}{Lee
  et~al\mbox{.}}{2013}]%
        {lee2013internet}
\bibfield{author}{\bibinfo{person}{Gyu~Myoung Lee}, \bibinfo{person}{Noel
  Crespi}, \bibinfo{person}{Jun~Kyun Choi}, {and} \bibinfo{person}{Matthieu
  Boussard}.} \bibinfo{year}{2013}\natexlab{}.
\newblock \showarticletitle{Internet of things}.
\newblock In \bibinfo{booktitle}{\emph{Evolution of Telecommunication
  Services}}. \bibinfo{publisher}{Springer}, \bibinfo{pages}{257--282}.
\newblock


\bibitem[\protect\citeauthoryear{Norman}{Norman}{2013}]%
        {norman2013design}
\bibfield{author}{\bibinfo{person}{Don Norman}.}
  \bibinfo{year}{2013}\natexlab{}.
\newblock \bibinfo{booktitle}{\emph{The design of everyday things: Revised and
  expanded edition}}.
\newblock \bibinfo{publisher}{Constellation}.
\newblock


\bibitem[\protect\citeauthoryear{Ognyanova}{Ognyanova}{2017}]%
        {ognyanova_2018}
\bibfield{author}{\bibinfo{person}{Katya Ognyanova}.}
  \bibinfo{year}{2017}\natexlab{}.
\newblock \bibinfo{title}{Static and dynamic network visualization with R}.
\newblock
\newblock
\newblock
\shownote{\url{http://kateto.net/network-visualization/}.}


\bibitem[\protect\citeauthoryear{Ratnakar}{Ratnakar}{2007}]%
        {ratnakar2007smart}
\bibfield{author}{\bibinfo{person}{Nitesh Ratnakar}.}
  \bibinfo{year}{2007}\natexlab{}.
\newblock \bibinfo{title}{Smart medicine container}.
\newblock
\newblock
\newblock
\shownote{US Patent 7,269,476.}


\bibitem[\protect\citeauthoryear{Reckwitz}{Reckwitz}{2002}]%
        {reckwitz2002toward}
\bibfield{author}{\bibinfo{person}{Andreas Reckwitz}.}
  \bibinfo{year}{2002}\natexlab{}.
\newblock \showarticletitle{Toward a theory of social practices: A development
  in culturalist theorizing}.
\newblock \bibinfo{journal}{\emph{European journal of social theory}}
  \bibinfo{volume}{5}, \bibinfo{number}{2} (\bibinfo{year}{2002}),
  \bibinfo{pages}{243--263}.
\newblock


\bibitem[\protect\citeauthoryear{Schneider and Kroner}{Schneider and
  Kroner}{2008}]%
        {schneider2008smart}
\bibfield{author}{\bibinfo{person}{Michael Schneider} {and}
  \bibinfo{person}{Alexander Kroner}.} \bibinfo{year}{2008}\natexlab{}.
\newblock \showarticletitle{The smart pizza packing: An application of object
  memories}.
\newblock  (\bibinfo{year}{2008}).
\newblock


\end{thebibliography}

\begin{appendix}
 
\end{appendix}

\end{document}